\documentstyle[aps,prl,floats,twocolumn,epsfig]{revtex}
\begin{document}
\title{Charge and spin addition energies of one dimensional quantum dot.}
\author{T. Kleimann$^{a}$, M. Sassetti$^{a}$, B. Kramer$^{a*}$, 
and A. Yacoby$^{b}$}
\address{$^{a}$Dipartimento di Fisica, INFM, Universit{\`a} di Genova, 
  Via Dodecaneso 33, 16146 Genova, Italy\\
$^{b}$Dept. of Condensed Matter Physics, 
The Weizmann Institute of Science, Rehovot, 76100 Israel}
\author{{\small (June 30, 2000)}\vspace{3mm}}
\author{\parbox{14cm}
{\parindent4mm
\baselineskip11pt%
{\small We derive the effective action for a one dimensional electron
  island formed between a double barrier in a single channel quantum
  wire including the electron spin. Current and energy addition terms
  corresponding to charge and spin are identified. The influence of
  the range and the strength of the electron interaction and other
  system parameters on the charge and spin addition energies, and on
  the excitation spectra of the modes confined within the island is
  studied. We find by comparison with experiment that spin excitations
  in addition to non-zero range of the interaction and inhomogeneity
  effects are important for understanding the electron transport
  through one dimensional quantum islands in cleaved-edge-overgrowth
  systems.}
\vspace{4mm} } } 
\author{\parbox{14cm} {\small PACS
numbers: 71.10.Pm, 73.23.-b, 23.23.Hk } } 
\maketitle
\section{Introduction}

Recently, considerable progress in the science and technology of
semiconductor nanostructures has been made. The experimental
realization of one-dimensional (1D) quantum wires has opened new
routes to investigating the influences of interactions and impurities
on electron transport at low temperatures where quantum effects
dominate \cite{tar95,yac96,yac97}. In spite of numerous theoretical
results obtained for the Luttinger liquid model \cite{tom,lutt,hal},
which have been used for interpreting the data, complete physical
understanding of the experiments has not been achieved yet. It is
therefore useful to study microscopically the combined effects of
interactions and impurities on 1D quantum transport within more
realistic models and to compare the results with experimental data.

The Luttinger liquid model allows to take into account interactions
between Fermions in 1D exactly. The linear conductance of a clean,
infinitely long (spin degenerate) Luttinger liquid has been predicted
to be $G=2e^2g_{\rho}/h$, the universal conductance
renormalized by an interaction constant $g_{\rho}$ \cite{kane}.
However, DC transport experiments on quasi-1D semiconductor quantum
wires at relatively high temperatures showed only quantization in
terms of the universal conductance $G=2e^2/h$. This has been
interpreted either in terms of a compensating, self-consistent
renormalization of the external electric driving field
\cite{kawa,oreg,cun} or using non-interacting leads connected to the
quantum wire \cite{maslov,pono,safi} which eventually determine the
conductance. However, at lower temperatures, temperature-dependent
deviations from the universal conductance steps have been found
\begin{equation}
  \label{power}
 \delta G\propto T^{\kappa}
\end{equation}
($\kappa <0 $). This has been attributed to weak  scattering at the
random impurity potential in a Luttinger liquid \cite{maslov95}.

When the electron density in the GaAs/AlGaAs-quan\-tum wires
fabri\-cated by using the cleaved-edge-over\-growth technique is
decreased by applying a voltage to an external gate, eventually even
the lowest electronic subband can be depopulated \cite{aus99} and the
region of Coulomb blockade reached. Here, the mean electron density is
sufficiently low, such that only very few maxima of the random
potential of the impurities are higher than the Fermi level. A {\em 1D
  quantum island} can be formed between two potential maxima in such a
wire. Electron transport is then dominated by charging effects
\cite{averin,mei90,leok}. At temperatures lower than the charging
energy, the linear conductance shows discrete peaks corresponding to
transferring exactly one electron through the quantum island.

In this region, it has been detected that the temperature dependence
of the intrinsic width $\Gamma$ of several conductance peaks is
modified by the correlations between the electrons and, instead of
being independent of $T$, shows non-analytical power-law behaviour
($\lambda > 0$)
\begin{equation}
  \label{resonant}
  \Gamma(T) \propto T^{\lambda}\, .
\end{equation}
Such a behaviour has been predicted in the sequential tunneling regime
for resonant  transport through a   quantum dot coupled to  a spinless
Luttinger liquid \cite{furu98} and for a double barrier in a Luttinger
liquid \cite{furu93,braggio}.

In non-linear transport, collective excited states of the electrons in
the island can contribute \cite{johns,weis}. The low-temperature
current-voltage characteristics shows fine structure within the
``Coulomb steps'' which reflects the excitation spectrum of the
electrons in the quantum dot. Especially in 1D, strong effects of the
spin of the correlated electrons have been predicted by using a
sequential tunneling model combined with the exact spectrum of
eigenvalues and the corresponding states \cite{weinmann}. In the above
mentioned experiments \cite{aus99}, contributions of the excited
states in the 1D island have been found. In addition, one of the
corresponding peaks in the differential non-linear conductance showed
an interaction-induced non-analytical temperature dependence
consistent with the above Luttinger liquid behaviour (\ref{resonant}).
Thus, correlations beyond the phenomenological charging model seem to
be present in these semiconductor quantum wires. This is also
indicated by recent results of Raman scattering experiments, which
strongly suggest that semiconductor quantum wires are very probably
dominated by non-Fermi liquid behaviour \cite{sk98}. In view of the
existing broad theoretical understanding of 1D non-Fermi liquids
\cite{voit}, it is highly desirable to provide quantitative results
for charging and correlation phenomena in these systems by using
reasonably realistic and controllable models, including non-zero range
interaction between the electrons as well as the potential of the
impurities.

Efforts in this direction have been made by studying transport through
a single barrier in a Luttinger liquid including long
range interactions \cite{fabri94,funa94,csk97,sass97}. Charging in the
presence of two impurities has been studied for spinless electrons
with interaction range much smaller than the distance between the two
impurities \cite{egger}. The crossover from a Luttinger liquid with
long-range interaction to a Wigner crystal in the presence of two
symmetrically arranged potential barriers has been investigated
\cite{mau}.

In the present paper, we establish the fundamentals of a transport
theory for 1D quantum dots embedded in a 1D electron system of
correlated electrons {\em with spin} which turns out to be
considerably different from the spinless case. Especially, we
concentrate here on the energies for charge and spin additions to the
electron island formed between two strong impurities in a
single-channel quantum wire of diameter $d$ with an electron
interaction of non-zero range. The quantum wire is described by using
the Luttinger liquid model with spin \cite{voit}. The bias electric
field is assumed to have an arbitrary spatial shape. The interaction
is assumed as a 3D Coulomb potential, which we project to 1D by using
the Gaussian wave function associated with the lowest state in a
parabolic, cylindrically symmetric confinement potential.  Screening
is introduced by an infinite metallic plane at a distance $D$ parallel
to the wire.  For comparison, we comment also on results obtained for
a projected screened 3D Coulomb interaction.

With the imaginary-time path-integral method, we determine the
effective action in which charging and transport contributions are
identified. The strength and the range of the interaction, $V_{0}$ and
$D$, respectively, and the length of the 1D quantum island -- the
distance between two delta-function-like potential barriers $a$ --
determine characteristic energy scales related to the charge and spin
excitations that appear in the model. We find that the characteristic
energy related to the addition of charge to the island -- {\em the
  charging energy} -- always increases with increasing interaction
range until it saturates when the latter exceeds considerably the
length of the island. On the other hand, the energy related to the
addition of a spin -- {\em the spin addition energy} -- is
approximately the same as that of non-interacting particles induced by
the Pauli principle. This is due to the smallness of the
(short-ranged) exchange interaction.

We observe that the spatial shape of the electric field enters
the effective action. The corresponding term can be interpreted as
simulating the influence of the voltage at a gate electrode in
experimental realizations.

Using our results, we have analyzed recent experimental data
\cite{aus99}. We attempted to deduce the parameters of the quantum
wire used in the experiment quantitatively. We found that it is not
possible to identify a region of parameters which is consistent with
{\em all} of the experimental findings. We conclude that depending on
the quantity considered, the frequency spectra of the collective
excitations or the temperature dependence of the peaks in the
differential conductance, experiments probe the interaction in
different regions of the 1D sample, namely within the island itself or
within the whole quantum wire extending along the edge of the sample.
This indicates that, together with spin and finite interaction range,
inhomogeneity effects have to be taken into consideration for
understanding the quantum transport in these 1D semiconductor systems.

The paper is organized as follows. Section \ref{model} briefly
describes the model. In Section \ref{action}, the effective action is
provided and discussed. In Section \ref{scurrent} we comment on the
conditions current transport. Quantitative results for the charging
energy are given in Section \ref{chargingenergy}.  Finally, we analyze
experimental data in Section \ref{discussion}.

\section{The Model}
\label{model}
\subsection{The Hamiltonian}

We use the bosonization technique \cite{hal,voit} for interacting
electrons with spin in 1D. The quantum dot is described by a double
barrier consisting of two delta-function potentials $V_{i}\,\delta
(x-x_{i})$ at $x_i\, (i=1,2)$ where $x_1<x_2$. An external electric
field ${\cal E}(x,t)=-\partial_{x} U(x,t)$ is assumed to induce
transport. The Hamiltonian is
\begin{equation}
  \label{hamiltonian}
 H= H_{0} + H_{\rm B} + H_{U}\,.  
\end{equation}

The first term describes the interacting electrons within the
Luttinger liquid model \cite{tom,lutt},
\begin{eqnarray}
\label{eq:1}
 H_{0} &=& 
\frac{\hbar v_{\rm F}}{2}\int {\rm d}x \Big[\Pi^{2}_{\rho}(x) 
+ (\partial_{x}\vartheta_{\rho}(x))^{2}\Big]\nonumber\\
&&\nonumber\\
&&+ \frac{1}{\pi}\int{\rm d}x\int{\rm d}x'\,
\partial_{x}\vartheta_{\rho}(x)\,V(x-x')\,
\partial_{x'}\vartheta_{\rho}(x')\nonumber\\
&&\nonumber\\
&&+ \frac{\hbar v_{\sigma}g_{\sigma}}{2}
\int {\rm d}x \left[\Pi_{\sigma}^{2}(x)
+ \frac{1}{g_{\sigma}^{2}}
(\partial_{x}\vartheta_{\sigma}(x))^{2}\right]\,,
\end{eqnarray}
where $v_{\sigma}$ is defined below and
\begin{equation}
  \label{gsigma}
  g_{\sigma}=\left(\frac{1+\eta_{\rm ex}}
               {1-\eta_{\rm ex}}\right)^{1/2}\,,
\end{equation}
with the exchange interaction matrix element $\eta_{\rm
  ex}\equiv\hat{V}(2k_{\rm F})/2\pi\hbar v_{\rm F}$.  In
Eq.~(\ref{eq:1}), the electrons are represented by conjugate bosonic
fields $\Pi_{\rho}, \vartheta_{\rho}$, and
$\Pi_{\sigma},\vartheta_{\sigma}$ associated with the collective
charge and spin density excitations, respectively. The system length
is $L$ and the Fermi velocity $v_{\rm F}$. The interaction, $V(x-x')$,
is a projection of a modified 3D Coulomb interaction onto the
$x$-direction (see below). It has the Fourier transform $\hat{V}(q)$
which is the dominant quantity in the dispersion relation of the
charge excitations \cite{s93}
\begin{equation}
  \label{eq:2}
\omega_{\rho}(q) = v_{\rm F}
\left|q\right|\left[1-\eta_{\rm ex}^2+
(1+\eta_{\rm ex})\frac{2\hat{V}(q)}{\pi\hbar v_{\rm F}} \right]^{1/2} 
\equiv v_{\rho}(q) |q|.
\end{equation}
The dispersion relation of the spin excitations,
\begin{equation}
  \label{eq:3}
\omega_{\sigma}(q)=v_{\rm F}|q|\left[1-
\eta_{\rm ex}^{2}
\right]^{1/2} \equiv v_{\sigma}|q|, 
\end{equation}
contains as the dominant part the exchange interaction which is
generally very small compared with $\hat{V}(q)$ at small $q$
\cite{sk98,s93}. This implies that the group velocity of the spin
excitations, $v_{\sigma}$ is very close to the Fermi velocity,
$g_{\sigma}\approx 1$. In the following, $\eta_{\rm ex}$ will therefore
be omitted whenever it is only a small correction.  

The contribution of the two localized impurities is
\begin{eqnarray}
H_{\rm B}&=& \rho_0\sum_{i=1,2} V_i \cos{[2k_{\rm F}x_{i}+
  \sqrt{2\pi}\vartheta_{\rho}(x_{i})]}\nonumber\\
  &&\qquad\qquad\times\cos{[\sqrt{2\pi}\vartheta_{\sigma}(x_{i})]} 
  \label{barrier}
\end{eqnarray}
where $\rho_0=2k_{\rm F}/\pi$ is the mean electron density.
Equation~(\ref{barrier}) is the potential energy of the impurities
corresponding to the number density of charges
\begin{eqnarray}
  \label{dens}
  &&\rho(x)=\rho_{\uparrow}(x)+\rho_{\downarrow}(x)
  =\rho_0+
   \sqrt{\frac{2}{\pi}}\,\partial_{x}\vartheta_{\rho}(x)\nonumber\\
&&\nonumber\\
  &&
 \qquad +\rho_0\cos{[2k_{\rm F}x +
    \sqrt{2\pi}\vartheta_{\rho}(x)]}\,
  \cos{[\sqrt{2\pi}\vartheta_{\sigma}(x)]}.
\end{eqnarray}
The second term in Eq.~(\ref{dens}) accounts for the slowly varying
part of the charge fluctuations. The third term represents the charge
density wave involved in the $2k_{\rm F}$ backscattering interference
between left and right moving electrons. It also couples the charge
with the long wave length part of the density of spins,
\begin{equation}
 \label{spin}
\rho_{\uparrow}(x)-\rho_{\downarrow}(x)\approx   
 \sqrt{\frac{2}{\pi}}\;\partial_{x}\vartheta_{\sigma}(x)\,,
\end{equation}
which is considered here with respect to a zero mean value. When
calculating the energy of the impurities, one obtains also a
contribution due to the second term in (\ref{dens}), besides the above
$H_{\rm B}$. As this represents a forward scattering process, it can
be eliminated by a unitary transformation and will not be considered
any further.

On the other hand, if a potential is slowly varying on the scale
$k_{\rm F}^{-1}$ its dominant contribution is mainly due to the
long-wave length part of the density (\ref{dens}). We assume this to
be the case for the bias electric field. The corresponding term in
the Hamiltonian is, with the elementary charge $e(>0)$
\begin{equation}
H_{U} = -e \sqrt{\frac{2}{\pi}} \,
       \int{\rm d}x \, U(x,t) 
       \partial_{x}\vartheta_{\rho}(x).
\label{drive}
\end{equation}

The presence of the two localized impurity naturally separates the
charge and spin degrees of freedom at ``bulk'' positions $x\neq
x_1,x_2$ from those at the barriers. It is useful to introduce
symmetric and antisymmetric variables for particle ($\nu = \rho$) and
spin densities ($\nu=\sigma$)
\begin{equation}
\label{variable}
N_{\nu}^{\pm} = \sqrt{\frac{2}{\pi}}\Big[\vartheta_{\nu}(x_{2}) \pm
\vartheta_{\nu}(x_{1})\Big]\, .
\end{equation}
The quantity $N_{\rho}^{-}$ is associated with the fluctuations of the
particle number within the island as compared to the mean particle
number $n_{0}=\rho_{0}(x_{2}-x_{1})$. The corresponding excess charge
is $Q = -eN_{\rho}^{-}$; $N_{\sigma}^{-}$ represents the $z$-component
of the fluctuation of the number of spins within the island
corresponding to a change of spin $N_{\sigma}^{-}/2$. The numbers of
imbalanced particles and spins between left and right leads are
represented by $N_{\nu}^{+}$. The DC current-voltage characteristic
can then be evaluated by considering the stationary limit of the
charge transfer through the dot in the presence of an external voltage
\begin{equation}
\label{current}
I=\frac{e}{2}\, \lim_{t\to\infty}
\langle \dot{N}_{\rho}^{+}(t)\rangle\, .
\end{equation}
The brackets $\langle\cdots\rangle$ include a thermal average over the
collective excitations at $x\neq x_{1},x_{2}$, and a statistical
average performed with the reduced density matrix for the degrees of
freedom at $x=x_{1},x_{2}$.

\subsection{The Interaction Energy}

We consider in the following two models which can be used for
discussing  experimental results quantitatively.

\subsubsection{Model 1.}

Because of gates and surrounding charges in the experiments done on
quantum wire, screening is always present for the interaction between
the electrons which normally interact via the 3D Coulomb interaction.
Although the geometry of the gates and the wire are certainly more
difficult in the experiment \cite{aus99}, we assume in the following
that we can describe the screening as being solely due to an infinite
metallic plane at a distance $D$ in the $y$-direction parallel to the
$(x,z)$-plane. The quantum wire is assumed in the $x$-direction. The
effective electron-electron interaction energy can then be calculated
by using the method of image charges
\begin{eqnarray}
  \label{image}
 & &V(\vec{r}-\vec{r'}) =
  \frac{V_0}{|\vec{r}-\vec{r'}|}\nonumber\\
  &&\quad- \frac{V_0}{\sqrt{(x-x')^{2}+(z-z')^{2}+(y+y'-2D)^{2}}},
\end{eqnarray}
with $V_{0}=e^{2}/4\pi\epsilon_{0}\epsilon$ and the dielectric
constant $\varepsilon_{0}\epsilon$.

We assume for simplicity a parabolic confinement of the
electrons perpendicular to the wire. The effective interaction in the
lowest subband can then be obtained from (\ref{image}) by projecting
with a normalized Gaussian confinement wave function corresponding to
the diameter $d$ of the wire. The Fourier transform of the resulting
effective interaction potential is for $D\gg d$
\begin{eqnarray}
\hat{V}(q)= 
V_0\left[\, {\rm e}^{d^2q^2/4} \,
{\rm E}_{1} \left(\frac{d^2q^2}{4}\right) - 
2{\rm K}_{0}(2Dq)\right]\,,
\label{1dqpot}  
\end{eqnarray}
where E$_{1}(z)$ is the exponential integral and ${\rm K}_{0}(z)$ the
modified Bessel function \cite{abramowitz}.  This expression shows
explicitely how the gate screens the Coulomb interaction. In the limit
$qd\to 0$, one obtains the finite value $\hat{V}(q\to 0)=
2V_0\left[\gamma/2+\log(2D/d)\right]$ ($\gamma= 0.57722$ Euler
constant). This implies a finite interaction constant
\begin{eqnarray}
\label{grho}
g_{\rho}&\equiv&\frac{v_{\rm F}}{v_{\rho}(q\to 0)}\nonumber\\
     &=&\left[1 + \eta \gamma +
       2\eta\log{\left(\frac{2D}{d}\right)}\right]^{-1/2}\,.
\end{eqnarray}
where $\eta= 2 V_0/\pi\hbar v_{\rm F}$.  The 1D equivalent of the
unscreened Coulomb interaction is obtained for $D \rightarrow \infty$.

\subsubsection{Model 2.}

We also consider the screened Coulomb interaction 
\begin{equation}
  \label{screenedpot}
  \hat{V}(\vec{r}-\vec{r'}) = 
  V_{0}\,\frac{{\rm e}^{-\alpha |\vec{r}-\vec{r'}|}}
  {|\vec{r}-\vec{r'}|}\,,  
\end{equation}
with a phenomenological screening length $\alpha^{-1}$. The
Fourier transform of its Gaussian projection to 1D is \cite{cun}
\begin{equation}
  \label{fscreenedpot}
  V(q)=V_{0}{\rm e}^{(d^2/4)(q^{2}+\alpha^{2})}
  {\rm E}_{1}\left(\frac{d^{2}}{4}[q^{2}+\alpha^{2}]\right),
\end{equation}
and the corresponding interaction constant
\begin{equation}
g_{\rho}=\left[1 -\eta\gamma + 
     2\eta\log{\left(\frac{2}{\alpha d}\right)}\right]^{-1/2}\,.
\end{equation}

\section{The Effective Action}
\label{action}

In order to evaluate the current-voltage characteristic one has to
perform a thermal average over the ``bulk modes'' at $x\neq
x_{1},x_{2}$. This can be done with the imaginary-time path
integral method \cite{sk}. The result of the integration is an
effective action $S_{\rm eff}$ which depends only on the four
variables defined in (\ref{variable}). In the continuous limit
($L\rightarrow \infty$), with the inverse temperature $\beta=1/k_{\rm
  B}T$ one obtains
\begin{eqnarray}
&& S_{\rm eff}[N_{\rho}^{\pm},N_{\sigma}^{\pm}]= 
\int_{0}^{\hbar \beta}  {\rm d}\tau \; 
H_{\rm B}[N_{\rho}^{\pm},N_{\sigma}^{\pm}]\nonumber\\
&&\nonumber\\
&&+\sum_{r=\pm}\sum_{\nu = \rho,\sigma}
\left[ \int_{0}^{\hbar \beta} 
\int_{0}^{\hbar \beta} {\rm d}\tau {\rm d}\tau' \;
N_{\nu}^{r}(\tau) K_{\nu}^{r}(\tau -\tau') 
N_{\nu}^{r}(\tau')\right. \nonumber \\
&&\nonumber\\
&&\left. \qquad\qquad- \delta_{\rho,\sigma}
\int_{0}^{\hbar \beta} {\rm d}\tau \;
N_{\rho}^{r}(\tau) {\cal L }^{r} (\tau) \right]\,.
\label{effaction}
\end{eqnarray}
The Fourier transforms, at Matsubara frequencies $\omega_n=2\pi
n/\hbar\beta$, of the dissipative kernels $K_{\nu}^{\pm}(\tau)$ and of
the effective ``forces'' ${\cal L}^{\pm}(\tau)$ are determined by the
dispersion relations of the collective modes (\ref{eq:2}) and
(\ref{eq:3}), respectively,
\begin{eqnarray}
&&\left[K_{\nu}^{\pm} (\omega_{n})\right]^{-1}= 
\frac{8 v_{\nu}g_{\nu}}{\hbar\pi^2} \int_{0}^{\infty}\!{\rm d}q  
\frac{1 \pm \cos [q(x_{1} -x_{2})]}{ \omega_{n}^2 +
\omega_{\nu}^2(q)}\,,\\
&&\nonumber\\
&&{\cal L}^{\pm}(\omega_{n})=  
\frac{4 e v_{\rm F}}{\hbar\pi^2} K^{\pm}_{\rho}(\omega_{n}) 
\int_{-\infty}^{\infty} {\rm d}x\; {\cal E}(x,\omega_{n})
\nonumber\\ 
&&\nonumber \\
&&\qquad\times \int_{0}^{\infty}\!{\rm d}q
\frac{\cos [q(x-x_{2}) ]\pm \cos [q(x-x_{1})]}{\omega_{n}^{2} 
+\omega_{\rho}^2(q)}\,.
\label{effdrv}
\end{eqnarray}

Both, $K_{\nu}^{\pm}$ and ${\cal L}^{\pm}$ contain the collective bulk
modes which introduce the interaction effects to be described below.
First of all, we note that $K_{\nu}^{+}(\omega_{n} \to 0)=0$
\cite{braggio,sk}. On the other hand, $K_{\nu}^{-}(\omega_{n}\to
0)\neq 0$. The latter describe the costs in energy for changing the
numbers of charges and/or spins on the island between the potential
barriers. The corresponding Euclidean action is
\begin{equation}
\label{costs}
S_{0}[N^{-}_{\rho},N^{-}_{\sigma}]=
\sum_{\nu=\rho, \sigma}
\frac{E_{\nu}}{2}  \int_{0}^{\hbar\beta}{\rm d}\tau\, 
(N_{\nu}^{-})^{2},    
\end{equation}
with the characteristic energies
\begin{equation}
\label{charging}
E_{\nu}= 2 K^{-}_{\nu}(\omega_n \to 0)
\qquad (\nu=\rho,\sigma).
\end{equation}
For $\nu=\rho$, this corresponds to the charging energy that is
supplied/gained, in order to transfer/remove one charge to/from the
island as compared with the mean value, $N_{\rho}^{-}=\pm 1$.
Correspondingly, for $\nu=\sigma$, the spin addition energy
$E_{\sigma}$ is needed/gained in order to change the spin by exactly
$\pm 1/2$. The Coulomb interaction in the dispersion
relation of the charge excitations, increases considerably the
charging energy $E_{\rho}$, in comparison with the spin addition
energy $E_{\sigma}$, which is only influenced by the (small)
exchange interaction. We always expect $E_{\rho} > E_{\sigma}$.

The frequency dependent parts of the kernels describe the dynamical
effects of the external leads and of the correlated excited states in
the dot. Their influence is described by spectral densities
$J^{\pm}_{\nu}(\omega)$ related via analytic continuation to the
imaginary-time kernels \cite{furu93,snw95}
\begin{equation}
  \label{spectral}
J^{\pm}_{\nu}(\omega)=
\frac{2}{\pi\hbar}{\rm Im}K^{\pm}_{\nu}(\omega_n\to+{\rm i}\omega)\,.
\end{equation}
Due to the non-zero range of the interaction,
analytic expressions for these densities are not available.
However, one can always extract their limits for $\omega\to 0$,
\begin{equation}
J^{\pm}_{\nu}(\omega\to 0)= \frac{A^{\pm}_{\nu}(g_{\nu})}{4g_{\nu}}
\omega,
\label{spectraldensity}
\end{equation}
where $A^{-}_{\rho}=g_{\rho}^{4}(E_{\rho}/E_{0})^{2}$ ($E_{0}=\hbar
\pi v_{\rm F}/2a$), and for the three other combinations of indices
$A^{\pm}_{\nu}= 1$. This limit describes the dissipative
influence of the low-frequency charge and spin excitations in the
external leads, $x < x_{1}$ and $x > x_{2}$. It holds also for finite
frequencies. However, these must be smaller than the frequency scale
corresponding to the range of the interaction, and smaller than the
characteristic excitation energy of the correlated electrons in the
dot.

Let us now discuss the driving forces. In general, ${\cal
  L}^{\pm}(\tau)$ depend in a quite complicated way on the dispersion
of the collective modes and on the shape of the electric field. We
focus on the DC limit where it is sufficient to evaluate the Fourier
components for $\omega_n\to 0$. In this case, the quantity ${\cal
  L}^{+}(\tau)$, which acts on the total transmitted charge, depends
only on the integral of the time independent electric field over the
entire system, the source-drain voltage $U\equiv
\int_{-\infty}^{\infty}{\rm d}x\,{\cal E}(x)$,
\begin{equation}
  \label{voltage}
  {\cal L}^{+}(\tau) = \frac{e U}{2}\,. 
\end{equation}
Since ${\cal L}^{+}$ is the part of the effective force that generates
the current transport, this result generalizes the one obtained
previously for only one impurity \cite{sk}. It can be easily derived
from Eq.~(\ref{effdrv}) by using the relation
\begin{eqnarray}
  \label{relation}
&&\frac{e^2v_{\rm F}}{\hbar\pi^2} \int_{0}^{\infty}{\rm d}q\,
\frac{\omega_n(1\pm\cos{qx})}
  {\omega_n^{2}+\omega_{\rho}^{2}(q)}
  \nonumber\\
\nonumber\\
&&\qquad\qquad\qquad
=\sigma_{0}(0,\omega_n)
    \pm\sigma_{0}(x,\omega_n)\,.
\end{eqnarray}
Here, $\sigma_{0}(x,\omega_n)$ is the frequency dependent non-local
conductivity of the Luttinger liquid per spin channel \cite{sk}, with
the DC limit $\sigma_0(x,0)=g_{\rho} e^2/h$.

On the other hand, ${\cal L}^{-}(\tau)$ acts on the excess charge on
the island, it does {\em not} generate a current. It depends on the
spatial shape of the electric field and can formally be written in
terms of the total charge $Q_{\cal E}$ accumulated between the points
$x_1$ and $x_2$ in the absence of the barriers as a consequence of
the presence of the DC electric field
\begin{equation}
\label{lminus}
{\cal L}^{-}(\tau)=\frac{E_{\rho} Q_{\cal E}}{e}\,,
\end{equation}
where  the charge is given by 
\begin{eqnarray}
Q_{\cal E}&&=2 \int_{-\infty}^{\infty}{\rm d}x'\,
{\cal E}(x')\nonumber\\ 
&&\times\lim_{\omega\to 0}
\left[\frac{\sigma_{0}(x_1-x',-{\rm i}\omega)- 
\sigma_{0}(x_2-x',-{\rm i}\omega)}
{{\rm i} \omega}\right]\,.
\end{eqnarray}

Equivalently, this can also be understood in terms of addition
energies. By introducing explicitly in Eqs.~(\ref{effdrv},
\ref{charging}) the dependence on the interval considered when
evaluating the addition energies, one finds
\begin{eqnarray}
  \label{lcharging}
   {\cal L}^{-}(\tau)&&= \frac{e}{2} E_{\rho}(x_{1}-x_{2})
\nonumber\\
   &&\times\int_{-\infty}^{\infty}{\rm d}x\,
   {\cal E}(x)\,\left[\frac{1}{E_{\rho}(x-x_1)}
    - \frac{1}{E_{\rho}(x-x_2)}\right].
\end{eqnarray}

It is reasonable to assume $x_{2,1}=\pm a/2$. If the
effective electric field has inversion symmetry, ${\cal L}^{-}$
vanishes. Without inversion symmetry, the electric field generates an
effective charge on the island which will influence the total current
via coupling between $N_{\rho}^{+}$ and $N_{\rho}^{-}$ due to the
impurity Hamiltonian $H_{\rm B}$. Physically, this externally
induced charge may be thought of as being generated by a voltage
$V_{\rm G}$ applied to an external gate which electrostatically
influences the charge on the island. Thus, the above
Eq.~(\ref{lminus}) can be interpreted as a term representing the
effect of the gate voltage in the phenomenological theory of Coulomb
blockade.

\section{Conditions for Current Transport}
\label{scurrent}

So far, we have discussed the influence of the spin and charge bulk
modes on the four variables $N^{\pm}_{\nu}$ that describe the quantum
dot.  In order to calculate explicitly the electrical current one has
to solve the equations of motion for the $N^{\pm}_{\nu}$.  This is
beyond the scope of the present work and will be discussed elsewhere.
Here, we briefly describe the results that are needed in the
following.

For barriers much higher then the charging energy, $V_{i}\gg
E_{\rho}$, the dynamics is dominated by tunneling events connecting
 the minima of $H_{\rm B}$ in the 4D $(N_{\rho}^{+}, N_{\rho}^{-},
N_{\sigma}^{+}, N_{\sigma}^{-})$-space \cite{kane}. For equal
barriers, $V_1=V_2=V$, the impurity Hamiltonian is
\begin{eqnarray}
H_{\rm B}&&= \rho_{0}V\left[
\cos\frac{\pi N^{+}_{\rho}}{2}\cos\frac{\pi N^{+}_{\sigma}}{2}
\cos\frac{\pi(n_0+ N^{-}_{\rho})}{2}\cos\frac{\pi N^{-}_{\sigma}}{2} 
\right.\nonumber\\
&&\nonumber\\
&&\left. +
\sin\frac{\pi N^{+}_{\rho}}{2}\sin\frac{\pi N^{+}_{\sigma}}{2}
\sin\frac{\pi(n_0+ N^{-}_{\rho})}{2}\sin\frac{\pi N^{-}_{\sigma}}{2}
\right]\,.
\label{barrier1}
\end{eqnarray}

The transitions between the minima of this function of four variables
correspond to different physical processes of transferring electrons
from one side to the other of the quantum dot. At very low
temperature, the dominant processes transfer the electron coherently
through the dot. In particular, when the number of particles in the
island is an odd integer there is spin degeneracy, $N_{\sigma}^{-}=\pm
1$. The island acts as a localized magnetic impurity, similar as in
the Kondo effect\cite{kane}. This leads to a Kondo resonance in the
transport through the island \cite{furu95,yi98}.

On the other hand, if the temperature is higher than the tunneling
rate through the single barrier, the dominant processes are sequential
tunneling events\cite{furu98,braggio}. The transfer of charge occurs
via uncorrelated single-electron hops into, and out of the island,
associated with corresponding changes in the total spin. This is
precisely the regime that recently has become accessible by using
cleaved-edge-overgrowth quantum wires \cite{aus99}. In this region,
one has to consider minima which correspond to pairs of even or odd
$N_{\rho}^{-}$ and $N_{\sigma}^{-}$. The dominant transport processes
are those which connect minima via transitions $N_{\rho}^{-}\to
N_{\rho}^{-}\pm 1$ associated with changes of the spin
$N_{\sigma}^{-}\to N_{\sigma}^{-}\pm 1$. For each of these processes
also the external charge and the spin change by $N_{\nu}^{+}\to
N_{\nu}^{+} \pm 1$.

The degeneracy of these minima is lifted by the charge and spin
addition energies $E_{\rho}$ and $E_{\sigma}$ which force the system
to select favourable charge and spin states in the island. These
selections become essential at low temperatures, $k_{\rm B}T<E_{\nu}$,
when current can flow only under resonant conditions. The latter can
be achieved in experiment by tuning external parameters, like the
source-drain voltage or the gate voltage, in order to create
degenerate charge states in the island.

\subsection{Linear Transport}
\label{lineartransport}

In the linear regime, $U\to 0$ for $T=0$, starting with
the island occupied by $n$ electrons, we expect that another electron
can enter and leave only if the difference between the ground state
energies of $n+1$ and $n$ electrons is aligned with the chemical
potential of the external semi-infinite Luttinger systems. The ground
state of an even number of electrons in the 1D island has the total
spin 0. On the other hand, the ground state of an odd number of
electrons can be assumed to have the spin $N_{\sigma}^{-} = \pm 1$
\cite{weinmann,lm62}. This implies the resonance condition
\begin{equation}
  \label{condition}
{\cal U}\left(n+1,\pm s_{n+1}\right)-{\cal U}(n,\pm s_{n})=0
\end{equation}
with ${\cal U}(n,\pm s_{n})$ the ground state energies with $n$
particles and total spins $s_{n}=0$ ($n$ even) or $s_{n}=1/2$ ($n$
odd), respectively.

With the above charge, spin and external gate terms,
Eqs.~(\ref{charging}) and (\ref{lminus}), these conditions become
\begin{equation}
E_{\rho}\left(n-n_0-n_{\rm G}+\frac{1}{2}\right) + 
(-1)^{n} \frac{E_{\sigma}}{2}=0.
\end{equation}
The variable $n_{\rm G}= eV_{\rm G}\delta/E_{\rho}$ represents the
number of induced particles due to the coupling to a gate at which the
voltage $V_{\rm G}$ is applied, with a proportionality factor $\delta$
which can be determined experimentally. The zero of energy has been
assumed to be given by the external chemical potential in
Eq.~(\ref{condition}).

From the above expression one can see that the distance of the peaks
of the linear conductance when changing the gate voltage are given by
$\Delta V_{\rm G}=(E_{\rho}+(-1)^{n}E_{\sigma})/e\delta \approx
E_{\rho}/e\delta$ since $E_{\sigma}\ll E_{\rho}$. Having independent
information on $\delta$ one can extract the value of the charging
energy $E_{\rho}$ (in principle also for the spin addition energy
$E_{\sigma}$) from the experimental data.

In order to evaluate the current as a function of temperature and/or
source-drain voltage, one needs to consider the behavior of the
spectral densities given in Eq.~(\ref{spectral}). In the sequential
tunneling regime, one can show that only their sum enters the
transport \cite{braggio}
\begin{equation}
J(\omega)=\sum_{r=\pm}\sum_{\nu=\rho,\sigma}J^{r}_{\nu}(\omega)\, .
\end{equation}
The frequency behavior of this determines the current-voltage
characteristics both in the linear and in the non-linear regimes (see
below). In particular, it determines the exponent of the power-law
dependencies of the current as a function of temperature and/or the
bias voltage.
 
For temperatures lower than the excitation energy in the quantum dot,
the low-frequency behavior of the spectral density is mainly
determined by the charge and spin excitations in the leads. Thus, the
power-law dependence of the conductance peaks is dominated by the
interaction in the regions of the quantum wire {\em outside} of the
electron island.  From Eqs.~(\ref{spectral},\ref{spectraldensity}) one
obtains
\begin{equation}
\label{spectral1}
J(\omega)\approx  J_{\rm leads}(\omega)=
\frac{\omega}{2}\left(\frac{1}{g_{\sigma}}+
\frac{1+A^{-}_{\rho}}{2 g_{\rho}}\right)\,.
\end{equation}
Eq.~(\ref{spectral1}) generalizes the results obtained previously for
spinless electrons and zero-range interaction \cite{furu98,braggio},
where $J(\omega)\approx\omega/g_{\rho}$. We conclude that the presence
of the spin in the leads introduces an effective interaction strength
\begin{equation}
\label{geff}
\frac{1}{g_{\rm eff}}=\frac{1}{4}\left(\frac{1+A^{-}_{\rho}}{g_{\rho}}
+\frac{2}{g_{\sigma}}\right)\,,
\end{equation}
which determines the exponents of the power laws.

For example, it has been shown for spinless electrons
\cite{furu98,braggio} that the intrinsic width $\Gamma(T)$ of the
linear conductance Coulomb peak depends on the temperature. For low
temperatures this has been found to be given by $\Gamma(T)\propto
T^{1/g_{\rho}-1}$. Correspondingly, with spin, one finds instead
\begin{equation}
\label{gamma}
\Gamma(T)\propto  T^{1/g_{\rm eff}-1}\,.
\end{equation}

\subsection{Non-Linear Transport}
\label{nonlinear}

One can investigate the non-linear current-voltage characteristic by
increasing the source drain voltage $U$. In this case, the
current-voltage characteristic shows the Coulomb staircase associated
with transitions between successive ground states of the electrons in
the quantum dot. In addition, fine structure appears which reflects
the excitation spectra of the correlated electrons. They can be due to
collective charge and spin modes or to particular polarizations of the
spins, $N_{\sigma}^{-}\neq 0,\pm 1$. The former, for a voltage drop
$U$ at the potential barriers (fixing the chemical potential in one of
the leads to be equal to that in the dot), has a maximum periodicity
\begin{equation}
\label{staircase}
U=\mu_0(n+1)-\mu_0(n)=E_{\rho}+(-1)^{n+1}E_{\sigma}  
\end{equation}
where $\mu_0(n)$ is the electrochemical potential of $n$ electrons on
the quantum island.

In the present model, the possible particle-hole excitations are
collective spin and charge modes confined within the island. In a
completely isolated island, the corresponding energy spectrum would be
discrete, $\omega_{\rho}(q_{m})$ and $\omega_{\sigma}(q_{m})$, due to
the discretization of the wave number $q_{m}=\pi m/a$ associated with
the confinement. Due to the coupling to the electrons in the leads via
the interaction, these levels are broadened. In the following
argument, we assume that this broadening is negligible.

The screened Coulomb interaction causes a non-linear dispersion
relation for the charge modes in the infinite Luttinger system. This
leads to non-equidistant charge excitation energies in the quantum
dot,
\begin{equation}
\Delta\epsilon_{\rho}(q_{m})=\hbar 
\left[\omega_{\rho}(q_{m+1})-\omega_{\rho}(q_{m})\right]\,.
\end{equation}
Their explicit values depend on the ratio between the distance $a$ and
the range of the interaction, $D$ or $\alpha^{-1}$.

For values of $a$ much larger than this range, the first excited
charge modes are equidistant. They are given by the charge-sound
velocity for $q\to 0$, $v_{\rho}=v_{\rm F}/g_{\rho}$,
\begin{equation}
\label{excitation}
\Delta\epsilon_{\rho}=\frac{\hbar\pi v_{\rho}}{a}=
    \frac{\hbar\pi v_{\rm F}}{a g_{\rho}}\equiv 
    \frac{2E_{0}}{g_{\rho}}\,.
\end{equation}
In the opposite limit, the non linear dispersion is already affecting
strongly the first excitation. This results in a value smaller than
Eq.~(\ref{excitation}), for model~1,
\begin{equation}
  \label{longrange}
\Delta\epsilon_{\rho}= \frac{\hbar\pi v_{\rm F}}{a}
     \sqrt{1-\gamma\eta +
       2\eta\log{\left(\frac{2a}{\pi d}\right)}}
     \equiv \frac{2E_{0}}{f_{\rho}}\,. 
\end{equation}

For the spin excitations, the dispersion in the infinite system is
linear and the discrete spectrum is equidistant
\begin{equation}
\Delta\epsilon_{\sigma}=\frac{\hbar\pi v_{\sigma}}{a}
  \equiv 2E_{0}\frac{v_{\sigma}}{v_{\rm F}}
\end{equation}
with the spin mode velocity $v_{\sigma}(\approx v_{\rm F})$.

\section{The Charging Energy}
\label{chargingenergy}

The above discussion emphasizes the {\em importance of charge and spin
  addition energies}, and of the {\em charge and spin excitation
  energies} for the linear and non-linear transport properties. It is
therefore crucial to evaluate these quantities microscopically and
determine their dependencies on the model parameters, especially in
view of comparisons with the experimental data of \cite{aus99}.

We analyze now in more detail the charging and spin energies of the
Luttinger island as a function of the parameters of above Model~1. In
addition to the interaction strength $V_{0}$, which contains as an
essential quantity the dielectric constant $\varepsilon $, we have the
Fermi energy $E_{\rm F}$ and three geometrical parameters: the
distance $D$ between the 1D system and the infinite metallic plane,
the diameter $d$ of the wire and the length $a$ of the island.

The spin addition energy can be easily evaluated from the simple
dispersion relation Eq.~(\ref{eq:3})
\begin{equation}
  \label{spincharge}
  E_{\sigma}=\frac{\pi\hbar}{2a}\frac{v_{\sigma}}{g_{\sigma}}\,.
\end{equation}
It is the same as the addition energy of non-interacting particles
which is related to the Pauli principle and due to the
quasi-discretization of the spin energies in the dot.

\begin{figure}[htb]
\begin{center}
\setlength{\unitlength}{1.0cm}
\begin{picture}(8.4,5.5)
\epsfig{file=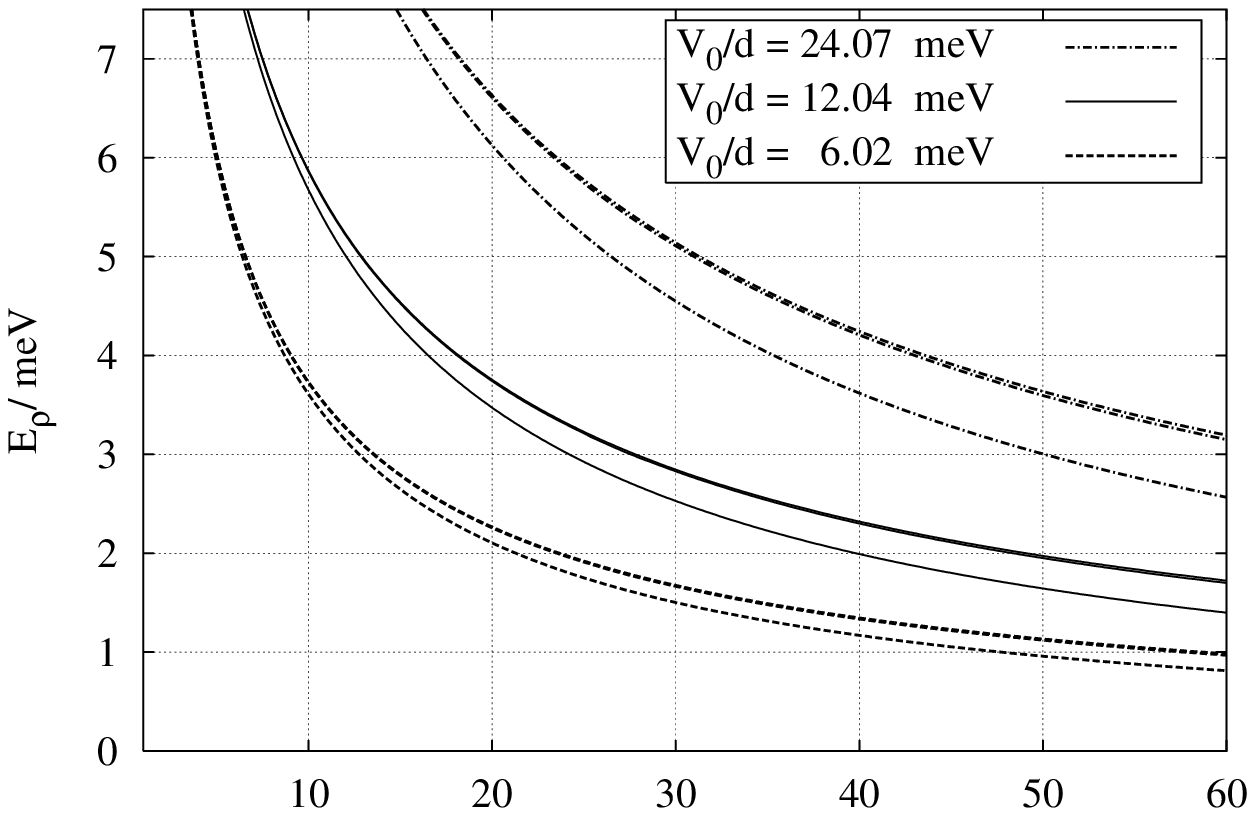,height=6.0cm}
\end{picture}
\begin{picture}(8.4,5.5)
\epsfig{file=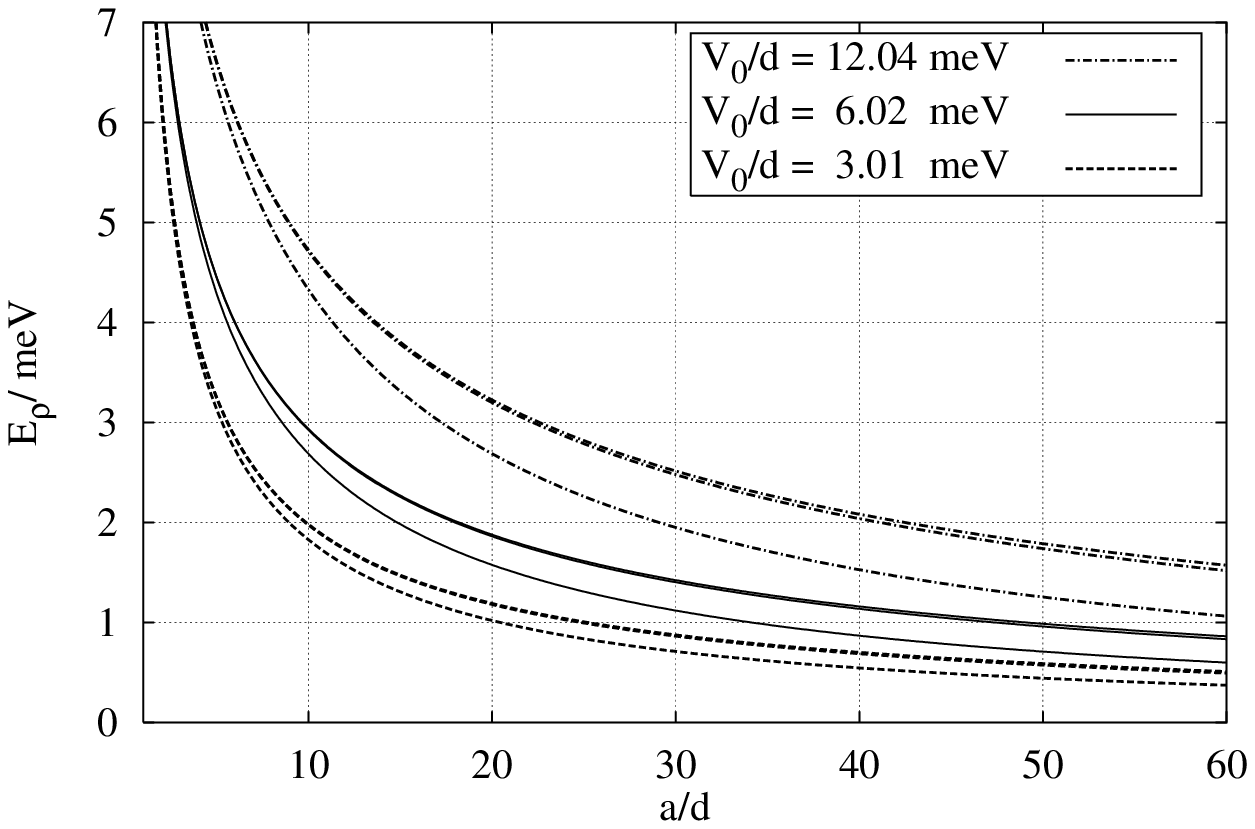,height=6.0cm}
\end{picture}
\caption{Charging energy $E_{\rho}$ in meV 
  for Model 1 as a function of $a/d$ (Fermi energy $E_{\rm F}=3$\,meV,
  effective electron mass $m=0.067m_{0}$). Top: $d=10$nm; interaction
  strengths $V_{0}/d=24.08$\,meV (dashed-dotted), 12.04\,meV (full
  lines,) 6.02\,meV (dotted), corresponding to $\epsilon = 6$, 12,
  24 (equivalent to $\eta=2V_{0}/\pi\hbar v_{\rm F}=2.04$, 1.02, 0.51,
  respectively), and $D/d=1000$, 100, 10 (top to bottom). Bottom:
  $d=20$nm; interaction strengths $V_{0}/d=12.04$\,meV
  (dashed-dotted), 6.02\,meV (full lines,) 3.01\,meV (dotted) and
  $D/d=500$, 50, 5 (top to bottom).}
\label{fig:1}
\end{center}
\end{figure}

The charging energy $E_{\rho}$ is evaluated numerically from
Eq.~(\ref{charging}) as a function of $a/d$. It is shown in
Fig.~\ref{fig:1} for different ratios $D/d$ and different interaction
parameters $V_{0}/d$ with $d=10$nm and $d=20$nm.  Different $V_{0}/d$
correspond to different $\epsilon$. Changing $E_{\rm F}$ between
2\,meV and 4\,meV, values that cover the experimentally relevant
region, changes the curves in Fig.~\ref{fig:1} only less than 10\%.
For small $a/d$, the charging energy diverges.  We consider only the
region $a/d>1$.  Asymptotically,
\begin{eqnarray}
\label{smallasympt}
\frac{E_{\rho}}{ E_{0}}&&\approx \left[1+\eta\gamma +
  2\eta\log{\left(\frac{2D}{d}\right)}\right]\equiv
\frac{1}{g_{\rho}^{2}}\quad (D\ll a)\,,\\
\nonumber\\
\frac{E_{\rho}}{E_{0}}&&\approx \left[1-\eta\gamma
  +2\eta\log{\left(\frac{2a}{\pi d}\right)}\right]
\equiv\frac{1}{f_{\rho}^2}\quad (D\gg a)\,.
\label{largeasy}
\end{eqnarray}
$E_{\rho}$ is only weakly influenced by $D/d$:
changing $D/d$ by a two orders of magnitude  changes
 $E_{\rho}$ only by about 30\%.
 
 The charging energy always increases with increasing interaction
 range $D$, because the cost of energy for putting additional
 electrons into the island increases. Interaction ranges much larger
 than $a$ do not change the charging energy because only the
 short-range part of the repulsion between the electrons contributes.
 Therefore, $E_{\rho}$ approaches the asymptotic value of
 Eq.~(\ref{largeasy}) (Fig.~\ref{fig:1}, curves for largest $D/d$).
 For strong Coulomb interaction ($\eta=2V_{0}/\pi\hbar v_{\rm F} \gg
 1$) and $a\gg d$, this is
\begin{equation}
E_{\rho}=\frac{e^2}{2\pi\epsilon\epsilon_0 a}
\log\left(\frac{2a}{d}\right)
\equiv \frac{e^2}{C}\,.
\end{equation}
Here, the $C$ is the classical self-capacitance of a cylinder of the
length $a$. The stronger the interaction (increasing $V_{0}$)
the larger is $E_{\rho}$, very similar to the behaviour
of a classical capacitor with a dielectric, $C\propto \varepsilon$.

Very similar results, with $d/D$ replaced by $\alpha d$, are obtained
by using Model~2. Basically, this tells that screening effects may
well be described by a global parameter, irrespective of the
underlying microscopic model.

As mentioned earlier, the non-linear transport shows Coulomb steps
with widths given approximately by $E_{\rho}$, and fine structure due
to the excited states. For a first estimate of the strength of the
interaction is very useful to estimate how many excited states are
present within the energy window given by $E_{\rho}$. Figure
\ref{fig:3} shows the ratios between the energy $E_{\rho}$ and the
first collective excited (spin or charge) state in the electron island
$\Delta \epsilon_{\nu}$ ($\nu=\sigma, \rho$) for Model~1 as a function
of the interaction constant $g_{\rho}$. The change in $g_{\rho}$ is
obtained changing the distance $D$ from the gate, $g_{\rho}\to 0$
corresponds to $D\to \infty$. The ratio
$E_{\rho}/\Delta\epsilon_{\rho}$ is always smaller than
$E_{\rho}/\Delta\epsilon_{\sigma}$.

We expect to observe within the window $E_{\rho}$ much more levels due
to spin excitations than due to charge excitations;
$E_{\rho}/\Delta\epsilon_{\nu}$ always increases when $g_{\rho}$
decreases (increasing $D$). It saturates for $D\gg a$,
\begin{equation}
\frac{E_{\rho}}{\Delta\epsilon_{\rho}}=
\frac{1}{2f_{\rho}}\,;\qquad
\frac{E_{\rho}}{\Delta\epsilon_{\sigma}}=\frac{v_{\rm F}}{v_{\sigma}}
\frac{1}{2f_{\rho}^{2}}\,.
\end{equation}
In the opposite limit, $D\ll a$, the behavior is
described by the asymptotic expressions
\begin{equation}
\frac{E_{\rho}}{\Delta\epsilon_{\rho}}=\frac{1}{2g_{\rho}}\,;\qquad
\frac{E_{\rho}}{\Delta\epsilon_{\sigma}}=\frac{v_{\rm F}}
{v_{\sigma}}\frac{1}{2g_{\rho}^2}\,.
\end{equation}

\section{Comparison with Experiment}
\label{discussion}

In this section, we compare our results with the experimental data
obtained in \cite{aus99}. In that work, results of the temperature
dependence of the intrinsic width of the conductance peaks in the
region of Coulomb blockade on quantum wires fabricated with the
cleaved-edge-overgrowth technique have been reported. The data have
been found to be consistent with the power law
\begin{equation}
  \label{peakwidth}
  \Gamma (T) \propto T^{1/g^{*}-1},
\end{equation}
with $g^{*}\approx 0.82$ and $g^{*}\approx 0.74$ for a peak closer to
the onset of the conductance and the next lower peak, respectively.
In addition, information about the excited energy levels of correlated
electrons have been obtained by measuring the non-linear current
voltage characteristics. A {\em minimum} of five excited levels have
been observed for a given electron number. Presumably, the number of
excited levels is even larger since the majority of the excited island
states is generally not clearly observed in non-linear transport due
to matrix element effects \cite{j96}.

The data have been analyzed previously by assuming that within the
quantum wire, a quantum dot has been accidentally formed between two
maxima of the random potential of impurities. The electron spin has
been neglected. The parameters estimated from the experimental setup
are: length of wire $L\approx 5\,\mu$m; length of the electron island
$a\approx 100-200$ nm; (non-spherical) diameter of the wire $d\approx
10-25$ nm; distance to the gate $D\approx 0.5\;\mu$m. The charging
energy, as estimated from the distance between the conductance peaks,
has been given as $E_{\rm C}\approx 2.2$ meV. Since the observed
conductance peaks are equidistant within 10\% we deduce that
$E_{\sigma}\ll E_{\rho}$. A rough estimate of the Fermi energy is
$E_{\rm F}\approx 3$\,meV. With these parameters, the value of the
interaction constant has been estimated as $g_{\rho}\approx 0.4$
\cite{aus99}, clearly inconsistent with the above $g^{*}$ determined
from the temperature dependence of the conductance peaks.

Using our above microscopic approach which includes the influence of
the electron spin and finite range of the interaction, we confirm that
there is a discrepancy, though the estimate for $g_{\rho}$ from the
temperature dependence of the conductance peaks seems to be slightly
improved.

First, we estimate the length of the island $a$ from the charging
energy in Fig. \ref{fig:1}. As $E_{\rho}$ is relatively insensitive
with respect to changes of $D/d$ and $E_{\rm F}$ in the experimentally
relevant region of parameters, we assume $D/d=50$ with $E_{\rm
  F}=3$meV and $d=20$nm. Furthermore, $V_{0}/d=6.02$\,meV for
$\epsilon=12$. With the above 2.2\,meV we find $a/d\approx 16$. This
is consistent with the value given in \cite{aus99} within the
uncertainties.

\begin{figure}[htb]
\begin{center}
\setlength{\unitlength}{1.0cm}
\begin{picture}(8.4,5.5)
\epsfig{file=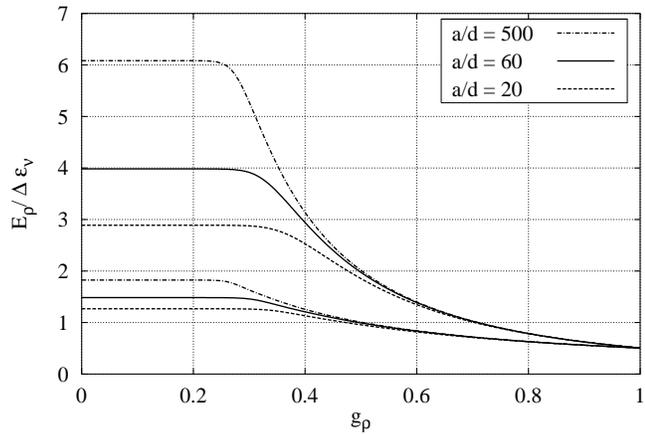,height=6.0cm}
\end{picture}
\caption{Ratios $E_{\rho}/\Delta\epsilon_{\nu}$ 
  ($\nu=\sigma, \rho$; upper/lower curves for fixed $a/d$) as a
  function of the interaction constant $g_{\rho}$ for $a/d=20$, 60,
  500 with $\epsilon =12$, $d=20$\,nm, $m=0.067m_{0}$, $g_{\sigma}=1$.
\label{fig:3}}
\end{center}
\end{figure}

By taking into account the spin and using an interaction of a finite
range, $g^{*}$ has to be identified with the effective interaction
$g_{\rm eff}$ in Eq.~(\ref{geff}). Since for the spin excitations
$g_{\sigma}\approx 1$, we find for $g^{*}=0.82$ and $0.74$ a
interaction constant $g_{\rho}^{*}\approx 0.6\pm 20\%$ by solving
Eq.~(\ref{geff}) for $E_{\rho}(g_{\rho})/E_{0}$ and comparing with
$E_{\rho}(g_{\rho})/2E_{0} = E_{\rho}/\Delta\epsilon_{\sigma}$ in
Fig.~\ref{fig:3}. This would imply that $D\approx d$
(Eq.~(\ref{grho})). On the other hand, by using Eq.~(\ref{grho}) with
the above parameters, especially $D\approx 50d$, we find a interaction
constant $g_{\rho} \approx 0.25\pm 20\%$ depending only weakly on
$D/d$. This is clearly {\em not} a quantitatively consistent result.
The discrepancy can be reduced by changing $\epsilon$. Also, taking
into account the exchange correction $\eta_{\rm ex}$ ($g_{\sigma}\neq
1$) gives some improvement. However, within reasonable values of all
of the parameters, one is in any case forced to conclude
$g_{\rho}^{*}\approx 2 g_{\rho}$.

Furthermore, with $g_{\rho}^{*} = 0.6$, one reads from
Fig.~\ref{fig:3} $E_{\rho}/\Delta\epsilon_{\rho}\approx
E_{\rho}/\Delta\epsilon_{\sigma}\approx 1$. Thus, about 1 to 2 excited
states corresponding to a given electron number should be observed in
non-linear transport spectroscopy. This is also {\em not} consistent
with the experiment. In order to observe a larger number of excited
states, the interaction constant must be considerably smaller,
$g_{\rho}<0.3$, such that the number of spin excited states within the
given window of $E_{\rho}$ is increased (cf. Fig. \ref{fig:3}). Taking
into account the exchange interaction ($v_{\sigma}<v_{\rm F}$) would
additionally decrease $\Delta\epsilon_{\sigma}$ (and $E_{\sigma}$)
increase $E_{\rho}$, and thus increase the number of excited states.

Despite the above inconsistency, the energetically lowest of the
excited states seen in the non-linear transport are predicted by our
approach to correspond to spin excitations. Without the latter there
is not at all any possibility to understand the results from the
non-linear transport even when using parameters that give consistent
results for the interaction constants. Given this scenario, only the
energetically highest experimentally observed state would correspond
to a charge excitation. The latter could be identified, for instance,
with the state denoted in \cite{aus99} as ``strongly coupled excited
state''.

One might conclude that the part of above quantitative discrepancy is
due to the Luttinger model being questionable for quantum wires in the
extreme low-density limit. However, one can also conclude that the
data obtained from non-linear transport {\em spectroscopy} have to be
interpreted by using a different interaction constant than that
obtained from the temperature dependence of the transport. This is
also supported by the theoretical derivation: the temperature
behaviour of the conductance peaks is dominated by the excitations in
the whole quantum wire, while the quasi-discrete excitation spectra
are showing the interaction in the region of the quantum island. We
are thus forced to conclude that {\em inhomogeneity} effects and the
influence of the contacts are an important issue for the understanding
of the correlations in the electron transport in these
cleaved-edge-overgrowth quantum wires, in addition to spin effects.

\section{Conclusion}
\label{conclusion}

In conclusion, we have derived the effective action for a quantum dot
formed by a double barrier potential with a realistic long range
interaction between the electrons, and, most importantly, including
the electron spin. We have identified an effective driving force
acting on the charge of the electrons, which turned out to be
independent of the shape of the driving electric field. Mass terms
originating from the dissipative degrees of freedom in the Luttinger
liquid have been found for both the charge {\em and the spin}. They
are identified with the charging energy in the case of the addition of
charge and with a spin addition energy for the spin. Also, an
effective force has been found which, in the DC limit, is only
non-zero when the driving electric field and/or the barriers are
asymmetric. This influences the transport via the coupling of modes at
the potential barriers. It can be attributed to charging of the
electron island via an external gate.

We have made an attempt to interpret quantitatively a recent
experiment which includes linear as well as non-linear transport data
in the region of Coulomb blockade. We find that even taking into
account the electron spin it is impossible to consistently fit {\em
  all} of the experimental data with the same interaction parameter.
It seems that different interaction strengths have to be used for
explaining the temperature dependence of the transport on the one
hand, and the excitation spectrum of the quantum dot on the other.

A possible conjecture to solve this puzzle is to assume that the
temperature behaviour of the conductance peaks is dominated by the
interaction among the electrons in the quantum wire along the edge of
the whole sample, in contrast to the excitation spectra which feel the
interaction strength near the quantum dot. This is qualitatively
consistent with the physical origin of the temperature dependence of
the conductance, namely the bulk modes outside the electron island.
This conjecture would also explain why $g_{\rho}^{*}\approx 0.6$, as
estimated from the temperature behaviour, is larger than that obtained
from the observed number of the excited states in the island,
$g_{\rho}< 0.3$. The interaction is weaker in the leads due to
screening induced by the contact regions with the attached 2D electron
gases while it can be assumed to be stronger within the island due to
the absence of nearby 2D electron gases.

The above interpretation also predicts that the energetically lowest
collective excitations in the quasi-1D quantum island observed in the
experiment are very likely spin excitations. The energies of the
lowest observed collective excited states are so small that it is hard
to believe that they can be related to charge modes since $v_{\rho} >
v_{\sigma}$, given the above estimates for the range and strength of
the interaction is of the correct order. This is consistent with a
previous suggestion based on a semi-phenomenological model in which
exact diagonalization and rate equations have been combined in order
to describe the non-linear transport through a 1D quantum island
containing only few electrons \cite{weinmann}.

\vspace*{0.25cm} \baselineskip=10pt{
\small 
\noindent 
This work has been initiated at the 225. 
International WE-Heraeus Seminar, October 11-15, 1999 in Bad Honnef,
and has been supported by the European Union within the TMR programme,
by the Deutsche Forschungsgemeinschaft within the SFB 508 of the
Universit\"at Hamburg, and by the Italian MURST via Cofinanziamento
98.}

\vspace*{0.25cm} \baselineskip=10pt{\small
\noindent $^{*}$ permanent address: I. Institut f\"ur Theoretische Physik,
Universit\"at Hamburg, Jungiusstra{\ss}e 9, 20355 Hamburg, Germany.}

%
%
\end{document}